\documentstyle[graphicx]{article}

\def\ie{\hbox{\sl i.e.\ }} \def\eg{\hbox{\sl e.g.\ }} \def\cf{\hbox{\sl 
c.f.\ }}

\newcommand{\cites}[1]{{\scriptsize$^{\cite{#1}}$}}

\begin{document}

\title{Note on the Kaplan--Yorke dimension and linear transport 
coefficients}

\author{Denis J. Evans,\footnote{Research School of Chemistry, The Australian 
National University, Canberra, ACT 0200, Australia} \ E. G. D. 
Cohen,\footnote{The Rockefeller University, 1230 York Avenue, New York, NY 
10021} \ Debra J. Searles,\footnote{Department of Chemistry, University of 
Queensland, Brisbane, Qld 4072, Australia} \ and F. 
Bonetto.\footnote{Department of Mathematics, Rutgers University, New 
Brunswick, NJ 08903}}

\maketitle

\begin{abstract} A number of relations between the Kaplan--Yorke dimension, 
phase space contraction, transport coefficients and the maximal Lyapunov 
exponents are given for dissipative thermostatted systems, subject to a 
small external field in a nonequilibrium stationary state.  A condition for 
the extensivity of phase space dimension reduction is given.  A new expression for 
the transport coefficients in terms of the Kaplan--Yorke dimension is 
derived.  Alternatively, the Kaplan--Yorke dimension for a dissipative 
macroscopic system can be expressed in terms of the transport coefficients 
of the system.  The agreement with computer simulations for an atomic fluid 
at small shear rates is very good.
\end{abstract}

\vspace{0.2cm} \noindent{Keywords: dynamical systems, KY--dimension, 
Lyapunov exponent, transport coefficient.} \vspace{0.2cm}

\section{INTRODUCTION}
\setcounter{equation}{1}

The Kaplan--Yorke (KY) or Lyapunov dimension was introduced\cites 
{FKYY83} as a conjecture relating the Hausdorff (H) dimension and the 
Lyapunov exponents of the invariant measure of a given dynamical 
system.\footnote {In the original paper only the natural invariant measure, 
\ie the weak limit of the Lebesgue measure or Sinai-Ruelle-Bowen measure, 
was considered.  In \cite{Y82} the result is formulated in a more general 
way such that it holds for every invariant measure.}\ This allows a 
computation of the H--dimension of the attractor of a dissipative dynamical 
system in terms of its Lyapunov exponents.  Its validity has been proven 
for two dimensional dynamical systems\cites{Y82} and for a rather large 
class of stochastic systems.\cites{LY88} In general one only knows that the 
KY--dimension is an upper bound for the H--dimension.\cites{LY85} Although 
it is not hard to construct rather artificial counter examples, it is 
generally believed that the conjecture holds for ``generic'' dynamical 
systems.\cites{R98}

In this paper we will discuss a relation (see Eq.  (\ref{main}) below) 
between the KY--dimension in large thermostatted systems and their physical 
properties such as the transport coefficients in a nonequilibrium 
stationary state.  Such a relation allows us to estimate the KY--dimension 
from a measurement of the transport coefficient.  In doing so we obtain a 
new relation between a dynamical quantity (the KY--dimension) and a 
physical quantity (the transport coefficient).

A difficulty in doing this is that while the dynamical quantities are 
usually defined for any finite number of particles $N$, the physical 
quantities usually refer to systems of very large $N$, so that one can 
meaningfully define intensive quantities, which only depend on intrinsic 
parameters like the number density $n=N/V$ rather than on $N$ and $V$ 
separately, where $V$ is the volume of the system.  We would like to 
state this by saying that strictly speaking, a thermodynamic limit has to 
be taken, \ie $N\to \infty$, $V\to \infty$ with the number density ($N/V\to 
n$) and other intensive quantities, such as in particular the shear rate, 
considered in this paper, held constant.  
This is straightforward if the linear transport coefficients are 
required and the limit of the external field $F_e \rightarrow 0$ is taken
before the limit $N \rightarrow \infty$, as in linear response theory.
However, for finite fields, changes in the behavior of the system can occur,
when the thermodynamic limit is approached - as e.g. the onset of turbulence
in a sheared system.\cites{H91}  In this paper we are interested only in the 
behavior of systems before such a transition takes place, like the laminar flow
of a sheared fluid considered in section 5. We think, 
nevertheless, that our results can be usefully formulated for
large systems using expressions like ``for sufficiently large $N$'' without 
taking the 
mathematical limit (see comment after Eq.  (\ref{sumlyapNc}) for a more precise 
discussion).  Although this expression is not mathematically well defined, we 
think that its meaning will be clear in any practical application (see note 
7 on page 14 for an attempt to clarify this point).  Moreover, because 
errors in intensive quantities are typically $O(N^{-1})$ where $N\sim 
10^{23}$, the approach is physically reasonable.

The above mentioned connection between a dynamical and 
thermodynamic treatment requires the usually discrete Lyapunov 
spectrum to be effectively replaced by a continuous intensive spectrum and 
an intrinsic version of the KY--dimension to be introduced.  In section 4 
we show how this can be implemented, after having introduced the basic 
equations which connect the dynamical and physical quantities in section 2, 
and deriving a new exact relation for the linear transport coefficients in 
section 3.

\section{BASIC RELATIONS AND DEFINITIONS}
\setcounter{equation}{0}

As has been shown before,\cites{ER85,HP88} there is a direct relationship 
between the sum of the non-zero Lyapunov exponents $ \lambda_i$ with 
$\lambda_i\ge\lambda_{i+1}$, $1\le i\le 2dN-f$ (where $N$ is the number of 
particles, $d$ the Cartesian dimension and $f$ the number of zero Lyapunov 
exponents) and the phase space contraction rate in a 
thermostatted system, subject to an external force $F_e$, in a non 
equilibrium stationary state, of the form:

\begin{equation}
\sum\limits_{i=1}^{2dN-f} \lambda _{i,N}(F_e)=\Lambda_N(F_e).
\label{sumlyap}\end{equation}
Here the subscript $N$ indicates the $N$-dependence of the various 
quantities in Eq.  (\ref{sumlyap}), $2dN-f$ is the effective number of degress 
of freedom in phase space of the system and $\Lambda_N(F_e)= {\partial \over 
\partial {\bf \Gamma}}\cdot\dot{{\bf \Gamma}}$ is the phase space 
contraction rate, where ${\bf \Gamma}$ stands for the collection 
of the coordinates and momenta of the $N$ particles and $\dot{{\bf 
\Gamma}}$ for its time derivative.  For macroscopic systems, \ie systems 
with very large $N$, one can use the equality of the dynamical phase space 
contraction and the physical entropy production\cites{HP88,CR98} and obtain 
from Eq.  (\ref{sumlyap}):

\begin{equation}
{1\over N}\sum\limits_{i=1}^{2dN-f}\lambda_{i,N}(F_e)= 
-{\sigma_N(F_e)\over 
nk_B}={J_N(F_e)F_e\over nk_BT}= -{L_N(F_e)F_e^2\over nk_BT}.
\label{sumlyapN}\end{equation}
Here $\sigma_N(F_e)$, $J_N(F_e)$ and $L_N(F_e)$ are the entropy production 
rate per unit volume, the dissipative flux and the transport coefficient 
respectively, induced in the system in the stationary state by the external 
force $F_e$, where a nonlinear constitutive relation 
$J_N(F_e)=-L_N(F_e)F_e$ has been used.  The subscript $N$ indicates the 
$N$-dependence for finite systems.  The kinetic temperature $T$ is 
determined by the relation:

\begin{equation}
{1\over dN-d-1}\sum\limits_{i=1}^{N} {{\bf p}_i^2 \over m}\equiv k_BT,
\label{ke}\end{equation}
where $m$ is the particle mass, $\{ {\bf p}_i,i=1,N\}$ are the peculiar 
momenta and $T$ is the kinetic temperature.

For systems at equilibrium the thermodynamic temperature appearing in 
Eq.  (\ref{ke}) should strictly only be calculated in the thermodynamic 
limit.  However, for nonequilibrium systems, as mentioned above 
the application of the limit $N\to\infty$ is not straightforward.

 Still, for sufficiently large values of $N$, Eq.  
(\ref{sumlyapN}) can be interpreted as:

\begin{equation}
{1\over N}\sum\limits_{i=1}^{2dN-f}\lambda_{i,N}(F_e)=
 -{\sigma(F_e)\over 
nk_B}+O(N^{-1})=-{L(F_e)F_e^2\over nk_BT}+O(N^{-1}),
\label{sumlyapNc}\end{equation}
where by $O(N^{-1})$ we mean that - at least at equilibrium - the finite size 
corrections can be bounded by a function of the form $CN^{-1}$ with $C$ of 
order 1.  Although this condition on the constant $C$ is not 
mathematically precisely defined, we discuss  
in section 5 numerical experiments, which will give an 
indication of the magnitude of the $O(N^{-1})$ corrections in the 
relationship between the KY--dimension and the viscosity. 
In what follows 
when we write $O(N^{-1})$, we will always intend it to carry the particular 
meaning given by the above discussion.

Since we are interested here in the linear ($F_e$) regime one can use that 
$L_N(F_e)=L_N+O(F_e^2)$,\footnote {Here and in what follows we assume that 
the transport coefficients are even in $F_e$.}\ where $L_N=L_N(0)\not=0$ is the 
linear transport coefficient.  We note that the left hand side of Eq.  
(\ref{sumlyapN}) is not restricted to small fields, due to its dynamical 
origin, and that the right hand side is related for small fields to the usual entropy 
production rate per unit volume of Irreversible Thermodynamics.\cites{GM84}

We now introduce the Kaplan--Yorke (KY) dimension.  If $N_{KY}$ is the 
largest integer for which $\sum\limits_{i=1}^{N_{KY}} 
\lambda_{i,N}(F_e)>0$, the KY--dimension, $D_{KY,N}$,\cites{FKYY83} for a 
finite system with a discrete Lyapunov spectrum, is defined 
by\cites{FKYY83,ER85}:

\begin{equation}
D_{KY,N}=N_{KY}+{\sum_{i=1}^{N_{KY}}\lambda_{i,N}(F_e)\over 
|\lambda_{N_{KY}+1,N}(F_e)|}\label{D_KY}\end{equation} 
where we have not indicated the $N$-dependence of $N_{KY}$.


\section{SMALL PHASE SPACE REDUCTION}
\setcounter{equation}{0}

In case the phase space dimension reduction is smaller than one, Eq.  
(\ref{D_KY}) can be reduced to the exact equation:

\begin{equation}\sum\limits_{i=1}^{2dN-f} \lambda_{i,N}(F_e)= 
\lambda_{min,N}(F_e)(2dN-f-D_{KY,N}(F_e))= {-\sigma_N(F_e)V\over 
k_B},\label{exa}\end{equation} 
where the minimum Lyapunov exponent, 
$\lambda_{min,N}(F_e)=\lambda_{2dN-f,N}(F_e)$.  From Eqs.  (\ref{exa}) and 
(\ref{sumlyapN}) we can then trivially calculate the linear (\ie the 
limiting zero field) transport coefficient in linear response theory as,

\begin{equation}L_N = \lim_{F_e\to 0} {(2dN-f-D_{KY,N}(F_e)) 
\lambda_{max,N}(F_e) nk_BT\over N F_e^2}.\label{tran}\end{equation} 
A similar relation has been obtained for the periodic Lorentz gas on the 
basis of periodic orbit theory.\cites{V92} We remark that from the point of 
view of linear response theory, \ie Eq. (\ref{tran}),  
 a phase space dimension reduction smaller 
than one occurs for any $N$, including $N \rightarrow \infty$.  In that 
case one can let $N \rightarrow \infty$ in Eq.  (\ref{tran}) and obtain a 
new exact relation for the linear transport coefficients, $L=\lim_{N\to 
\infty}L_N$, equivalent to the Green-Kubo formulae.  The corresponding 
expression for the KY--dimension of the steady state attractor for 
sufficiently small fields, is given by:

\begin{equation}D_{KY,N}(F_e)=2dN-f-{L F_e^2N\over \lambda_{max,N}n 
k_BT}+O(F_e^4)
\label{KY_e}.\end{equation}
To obtain the Eqs.  (\ref{tran}) and (\ref{KY_e}), we have used that for 
systems which are symplectic at equilibrium, (\ie all 
Hamiltonian equilibrium systems), one can write for small 
$F_e$: $\lambda_{max,N}(F_e)=\lambda_{max,N}+O(F_e^2)= 
-\lambda_{min,N}+O(F_e^2)$, where $\lambda_{max,N} \equiv \lambda_{max,N} 
(0)$ and $\lambda_{min,N} \equiv \lambda_{min,N}(0)$.  

For systems which satisfy the Conjugate Pairing 
Rule\cites{ECM90} (CPR) the sum of each conjugate pair, $i,i^*=2dN-f-i+1$ 
of Lyapunov exponents is

\begin{equation}\lambda_{i,N}(Fe)+\lambda_{i^*,N}(Fe)= {{-2\sigma_N(Fe)V} 
\over {k_B}(2dN-f)},\;\forall i.\label{CPR}\end{equation} 
We note that in nonequilibrium systems, the Conjugate Pairing Rule is 
expected to hold only in systems that are thermostatted homogeneously. In some systems there is
numerical evidence that the maximal exponents satisfy the Conjugate Pairing
Rule, while the other pairs do not (that is Eq. (\ref{CPR}) is true for
i=1).\cites{SEI98}  These systems are said to satisfy the weak Conjugate Pairing Rule (WCPR).  
By combining the Eqs.  
(\ref{exa}) and (\ref{CPR}), one obtains for sufficiently small fields:

\begin{equation}{D_{KY,N}(F_e)\over (dN-f/2)}=1-{\lambda_{max,N}(F_e) 
\over\lambda_{min,N}(F_e)}.\label{CPR2}\end{equation}

Substituting Eq.  (\ref{CPR}) into Eq.  (\ref{sumlyapN}) and using Eq.  
(\ref{KY_e}), one obtains another expression for the limiting 
KY--dimension, for sufficiently small fields:

\begin{equation}{D_{KY,N}(F_e) \over (dN-f/2)}=3+{\lambda_{min,N}(F_e) 
\over\lambda_{max,N}(F_e)}+O(F_e^4)\label{CPR3}\end{equation} 
In Eqs.  (\ref{CPR2}) and (\ref{CPR3}), we have chosen to use the maximal 
Lyapunov exponents as the conjugate pair in Eq.  (\ref{CPR}), and therefore 
these equations are valid provided WCPR is obeyed.  A similar 
looking formula, as Eq. (\ref{CPR3}) with the first two terms on the right hand side only, has 
been quoted in \cite{DGP95}.

As mentioned before, all the results in this section hold under the 
hypothesis that the phase space dimension reduction is smaller than unity.  
For this to be true for any given small $F_e$, $N$ is constricted to be of 
$O(F_e^{-2})$, or equivalently, for any given large $N,F_e$ to 
 be of 
$O(N^{-1/2})$, as can be seen from Eq.  (\ref{KY_e}), so that $F_e$ and $N$ 
are coupled.  \footnote {One might think that a phase space dimension reduction 
of one could 
hardly have any practical consequence in a macroscopic system whose phase 
space dimension is of the order of $10^{23}$.  As we will discuss in 
Section 6, such a very small phase space dimension reduction is expected to 
occur under physically realizable conditions.}

However, from a general physical point of view, we would like to have a 
theory for $D_{KY,N}$, which holds uniformly in $N$, \ie with Eq.  
(\ref{KY_e}) valid for every $N$ with a small but fixed $F_e$, so that $N$ 
and $F_e$ are independent variables.  Now, it is trivial to generalize 
Eq.  (\ref{exa}) to the case in which the dimensional reduction is greater 
than one. In fact one then obtains that:

\begin{eqnarray}
\sum\limits_{i=1}^{2dN-f} \lambda_{i,N}(F_e)&=& \lambda 
_{N_{KY}+1,N}(F_e)(2dN-f-D_{KY,N}(F_e)) \nonumber \\
& &+\sum\limits_{i=N_{KY}+2}^{2dN-f}(\lambda_{i,N}(F_e) - 
\lambda_{N_{KY}+1,N}(F_e)).
\label{exa_1}\end{eqnarray}
If we assume, as is usually done, that for sufficiently large $N$, 
$\lambda_{2dN-f-j}=\lambda _{2dN-f}+O(N^{-1})$ for fixed $j$ not varying with N; then
for any fixed dimensional reduction, Eq.  (\ref{exa_1}) simply 
becomes Eq.  (\ref{exa}) with a correction of $O(1)$ in $N$.  Keeping the
phase space dimension reduction fixed as $N$ increases still imposes a condition on $F_e$ of the
form discussed above. To better 
control the errors when $2dN-f-N_{KY}$ becomes large ($O(N)$), a more 
careful treatment of the Eq.  (\ref{exa}) is needed in order to obtain an 
expression uniform in $N$ for the phase space dimension reduction.

\section{LYAPUNOV SPECTRUM FOR VERY LARGE $N$ AND LARGE PHASE SPACE 
REDUCTION} \setcounter{equation}{0}

For large $N$ one can indeed reformulate the definition of $D_{KY,N}$ in a 
more analytical way.  Consider thereto the stepwise continuous 
function of a continuous variable $0<x\le 2dN-f$: 
$\tilde\lambda_N(x,F_e)=\lambda_{i,N}(F_e)$ for $i-1<x \le i$.  If one 
introduces the integral:

\begin{equation}\tilde\mu_N(x,F_e)=\int_0^x 
\tilde\lambda_N(y,F_e)dy,\label{mu}\end{equation} 
Eq.  (\ref{D_KY}) can be rewritten in the form:

\begin{equation}\tilde\mu_N(D_{KY,N}(F_e),F_e)=0.\label{fond}\end{equation}
Since $D_{KY,N}(F_e)$ as well as $\tilde\mu_N(x,F_e)$ 
are expected to be extensive quantities, \ie they are 
proportional to $N$ for large $N$, it is natural to define the quantity

\begin{equation}\delta_N(F_e)={D_{KY,N}(F_e)\over 
2dN-f}\label{delta}\end{equation}
where $\delta_N(F_e)$ is the dimension per effective degree of freedom 
($2dN-f$) in phase space.  
We now use again that, as $N$ grows, the difference 
$\lambda_{i,N}-\lambda_{i+1,N}$ is expected to go to zero as $N^{-1}$.  This suggests 
a possible rescaling of the variable $x$ in $\tilde\lambda_N(x,F_e)$ to 
define the function $\lambda_N(x,F_e)=\tilde\lambda_N(xN,F_e)$, \ie 
$\lambda_N(x,F_e)=\lambda_{i,N}(F_e)$ for ${i-1\over 2dN-f}<x\le {i\over 
2dN-f}$.  The function $\lambda_{i,N}(F_e)$ can then be expected to be well 
approximated by a continuous function of the variable $x$ when 
$N$ is sufficiently large.  Thus rewriting Eq.  
(\ref{mu}) as

\begin{equation}\mu_N(x,F_e)=\int_0^x\lambda_N(y,F_e)dy\label{mu1}\end{equation}
where $0 \le x \le 1$, Eq.  (\ref{fond}) is equivalent to:

\begin{equation}\mu_N(\delta_N(F_e),F_e)=0\label{fond1}\end{equation}
where $\delta_N(F_e)$ and $\mu_N(x,F_e)$ can be assumed to be intensive 
quantities.  We can make this more precise by the following crucial 
assumption on the nature of the function $\lambda_N(y,F_e)$, \ie of the 
Lyapunov spectrum:

\noindent{\bf Smoothness Hypothesis:} If $N$ is sufficiently large, one can 
write:

\begin{equation}\lambda_N(x,F_e)=l(x,F_e)+O(N^{-1})\label{hyp}\end{equation}
with $l(x,F_e)$ a smooth function\footnote {It is enough to assume that 
$l(x,F_e)\in L^1$ in $x$ and $C^2$ near $x=1$.  Moreover we will require 
that $l(x,F_e)$ is $C^2$ in $F_e$ for every $x$. This may not be the case
when a phase transition occurs.\cites{DP97}}\ of $x$ and $F_e$.

Eq.  (\ref{hyp}), as Eq.  (\ref{sumlyapNc}), is to be interpreted that for 
sufficiently large $N$, $|\lambda_N(x,F_e)-l(x,F_e)|<C'N^{-1}$ with $C'$ of 
order 1.  Similarly, as mentioned already above, all the $O(N^{-1})$ terms 
appearing in forthcoming equations (see \eg Eq.  (\ref{mainN})) have to be 
interpreted in this way and the constants obtained then can be directly 
expressed in term of $C'$.

\begin{figure}
\begin{center}
\includegraphics[scale=0.35]{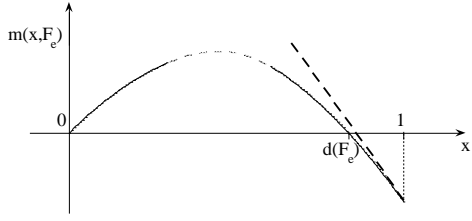}
\caption{Sketch of the integral of a continuous Lyapunov spectrum 
$m(x,F_e)$ for a dissipative system, starting with the largest Lyapunov 
exponent at $x=0$ and ending with the smallest Lyapunov exponent at $x=1$.  
The dashed line is the tangent at $x=1$.}
\label{Fig1}
\end{center}
\end{figure}

We now define, in analogy with Eqs. (\ref{mu1}) and (\ref{fond1}):

\begin{equation}m(x,F_e)=\int_0^xl(y,F_e)dy\label{m}\end{equation}
and $d(F_e)$ through the equation:

\begin{equation}m(d(F_e),F_e)=0\label{fond_m}\end{equation}
respectively.  Clearly our Hypothesis implies that:

\begin{equation}\delta_N(F_e)=d(F_e)+O(N^{-1})\label{hyp1}\end{equation}
where $d(F_e)$ and $m(F_e)$ are now intensive quantities.  The function 
$m(x,F_e)$ is sketched in Fig. 1 \ref{Fig1}, especially near $x=1$.

From this figure one easily deduces that:

\begin{equation}d(F_e)=1-{m(1,F_e)\over m'(1,F_e)}+O(m(1,F_e)^2)=1+{1\over 
l(1,F_e)} {\sigma(F_e)\over 2dnk_B}+O(F_e^4),\label{main}\end{equation} 
where $m'(F_e)$ is the derivative of $m(x,F_e)$ with respect to x at 
$x=1$.  
Here we used that $m'(1,F_e)=l(1,F_e)=\lambda_{min}(F_e)$ and that 
$m(1,F_e)$, is the sum over all Lyapunov exponents, divided by $(2dN-f)$, 
\ie the phase space contraction (or entropy production) rate per 
(effective) degree of freedom in phase space.\footnote {In a more analytic 
way Eq.  (\ref{main}) follows from the inverse function theorem together 
with our Smoothness Hypothesis and the fact that $m(0,0)=0$.}\ Using Eq.  
(\ref{sumlyapN}), Eq.  (\ref{main}) can be rewritten as:

\begin{equation}d(F_e)=1-{1\over \lambda_{max}}{LF_e^2\over 
2dnk_BT}+O(F_e^4)\label{main1}\end{equation} 
where the maximum and minimum Lyapunov exponents at equilibrium, ({\sl 
i.e.} when $F_e=0$) are $\lambda_{max}=l(0,0)$ and $\lambda_{min}=l(1,0)$, 
respectively. Moreover we have used that $\lambda_{max}=-\lambda_{min}$ 
and that $l(1,F_e)=l(1,0)+O(F_e^2)$.  In terms of the extensive 
quantity $D_{KY,N}(F_e)$, Eq.  (\ref{main1}) can be rewritten as:

\begin{equation}
{D_{KY,N}(F_e)\over 2dN-f}=1-{1\over \lambda_{max}}{LF_e^2\over 2dnk_BT}+ 
O(F_e^4)+O(N^{-1})\label{mainN}.\end{equation} 
where we kept the term $f/(2dN)$ of $O(N^{-1})$ on the left hand side of 
Eq. (\ref{mainN}) to facilitate comparison in figures 4 and 5 of section 5 for $N=32$ particles.
 We observe that Eq.  (\ref{mainN}) is 
formally very similar to Eq.  (\ref{KY_e}), except for the correction term 
$O(N^{-1})$ and the substitution of the asymptotic $\lambda_{max}$ for the 
finite $N$ value $\lambda_{max,N}$.  It is clear that the same 
generalization can be performed for the Eqs.  (\ref{CPR2}) and 
(\ref{CPR3}), if one notes that in the present context, the CPR, Eq.  
(\ref{CPR}), becomes:

\begin{equation}l(x,F_e)+l(1-x,F_e)= -{\sigma(F_e)\over dn 
k_B}.\label{pairN}\end{equation}

From Eq.  (\ref{main1}) it follows that in the linear regime the reduction 
in phase space dimension in large thermostatted dissipative systems is 
extensive.  For those systems for which the Smoothness Hypothesis holds, 
this result is exact.  The extensive nature of the reduction has been 
noted before.\cites{HP88,HP94,GC95,DHP98}

\section{NUMERICAL TEST}
\setcounter{equation}{0}

We tested our Smoothness Hypothesis as well as Eq.  
(\ref{mainN}) and equations derived from it assuming that the WCPR is valid, 
for a 
system of 32 WCA particles\cites{WCA71} undergoing shear flow in two 
dimensions.  Although this system does not satisfy the CPR, it does appear 
to satisfy WCPR to within $0.7\%$ when 
$N=32$\cites{SEI98} (which was within the limits of numerical error 
achieved).  The equations of motion for particles in such a system are the 
so-called SLLOD equations,\cites{EM90}

\begin{equation}\dot{\bf{q}}_i={\bf{p}}_i/m+{\bf{i}}\gamma y_i, \quad 
\dot{\bf{p}}_i={\bf{F}}_i-{\bf{i}}\gamma p_{yi}-\alpha 
{\bf{p}}_i.\label{sllod}\end{equation} 
Here, at not too large Reynolds numbers, the momenta, ${\bf{p}}_i$ are 
peculiar momenta, $\bf i$ is the unit vector in the $x$ direction and 
${\bf{F}}_i$ is the force exerted on particle $i$ by all the other 
particles, due to a Weeks-Chandler-Andersen pair interaction 
potential\cites{WCA71} between the particles.  The value of $\alpha$ 
is determined using Gauss' Principle of Least Constraint to keep the 
kinetic temperature fixed.\cites{EM90} The   
SLLOD equations of motion given by Eq.(\ref{sllod}) then model Couette flow 
when the Reynolds number is sufficiently small so that laminar flow is 
stable.\cites{EM90} \footnote {The $\alpha$ appearing in the Eq.  
(\ref{sllod}) is related to the phase space contraction rate $\Lambda$ in 
Eq.  (\ref{sumlyap}).  It is given by the relation $\sum^{2dN-f}_{i=1} 
\lambda_i =\Lambda= -dN\alpha+O(1)$.}

For this system the dissipative flux $J$ is just the xy element of the 
pressure tensor $P_{xy}$; the transport coefficient $L$ is the Newtonian 
shear viscosity $\eta$; and the external field $F_e$ is the shear rate 
$\gamma= \partial u_x/\partial y$,\cites{EM90} where $u_x$ is the local 
flow velocity in the $x$-direction in the system. The calculations were 
carried out at a reduced kinetic temperature of unity and a reduced density 
{\it{n}}$=N/V=0.8$.  All physical quantities occurring in the Eq.  
(\ref{sllod}) as well as the temperature and density are made dimensionless 
by reducing them with appropriate combinations of molecular quantities.  In 
particular the reduction factor for $\gamma$ amounts to about 1 ps$^{-1}$ 
for Argon.  (\cf \cite{HP94}) We note that for the SLLOD equations 
for shear flow in 2 dimensions using non autonomous Lees-Edwards periodic 
boundary conditions,\cites{EM90,PE98} $f=5$ (due to the conservation of 
kinetic energy, momentum and position of the center of mass).

Figures 2 and 3 represent a direct test of our Smoothness Hypothesis.  
Figure 2 gives the Lyapunov spectrum for a system at a shear rate 
$\gamma=0.05$ and for a number of particles $N=8,18,32$.  As can be easily 
observed the Lyapunov exponents for $N$=18 and 32 just fill the ``open 
spaces'' left by the exponents for $N$=8 and 18, respectively.

\begin{figure}
\begin{center}
\includegraphics[scale=0.35]{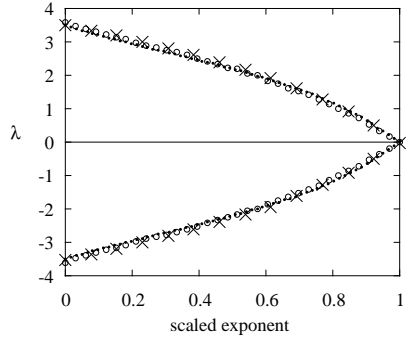}
\caption{Discrete Lyapunov spectra for the SLLOD Eqs.  
(\ref{sllod}) for $N=$8 (crosses), 18 (open circles) and 32 (small filled 
circles) and a shear rate $\gamma=0.05$, as a function of the scaled 
Lyapunov exponent pair index $(i-1)/(2N-2)$, which runs from 0 to 1 for $1 
\le i \le 2N-1$.}
\label{Fig2}
\end{center}
\end{figure}

\begin{figure}
\begin{center}
\includegraphics[scale=0.35]{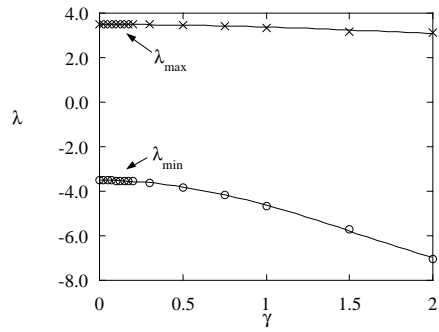}
\caption{Maximum and minimum Lyapunov exponents, 
$\lambda_{max}(\gamma)$ (crosses) and $\lambda_{min}(\gamma)$ (circles), 
respectively, as a function of $\gamma$.}
\label{Fig3}
\end{center}
\end{figure}

Figure 3 shows the behavior of $\lambda_{max}(\gamma)$ and 
$\lambda_{min}(\gamma)$ as functions of the externally applied field 
$\gamma$.  Although no numerical experiment can in general confirm a 
mathematical
hypothesis, the numerical results seem to 
agree very well with our Smoothness Hypothesis.

We now use the Lyapunov spectrum to compute $D_{KY,N}(\gamma)$.  This is 
shown in figure 4a where the $D_{KY,N}(\gamma)$ is plotted for $N=32$, 
$d=2$ and $0<\gamma<0.5$.  As can be easily seen, although from 
the definition Eq.  (\ref{D_KY}) one would generally expect to 
see discontinuities in the first derivatives of this function for those 
values of $\gamma$ for which $N_{KY}$ change (\cf \cite{ER1}), 
the function appears very smooth for a rather large range of values of 
$\gamma$.  This indicates that $N=32$ is already ``big enough'' 
to consider the Lyapunov spectrum as ``effectively continuous''.\footnote 
{More precisely one can say that for $N=32$ the numerical errors involved 
in computing the Lyapunov exponents are already larger than the errors 
introduced by neglecting the $O(N^{-1})$ correction.  This 
observation provides a more precise meaning of expressions like 
``sufficiently large $N$'' or ``a constant $C$ of order 1''.}\ This also 
permits us to check the validity of Eq.  (\ref{mainN}), and the large $N$ 
versions of Eqs.  (\ref{CPR2}) and (\ref{CPR3}), which in this case take 
the form:

\begin{eqnarray}
{D_{KY,N}(\gamma )\over 2dN-f}= 1-{\eta_N \gamma^2\over 
\lambda_{max,N} 2dnk_BT}+O(\gamma^4)+O(N^{-1})\label{sllod2}\\
{D_{KY,N}(\gamma )\over dN-f/2}= 1-{\lambda_{max,N}(\gamma )\over\lambda 
_{min,N}(\gamma )}+ O(\gamma^4)+O(N^{-1})\label{sllod3}\\
{D_{KY,N}(\gamma )\over dN-f/2}= 3+{\lambda _{min,N}(\gamma )\over \lambda 
_{max,N}(\gamma )}+O(\gamma^4)+O(N^{-1}),
\label{sllod4}
\end{eqnarray}
respectively.  Figure 4a shows that these expressions can  
describe the system studied over the range of fields considered and that 
the correction terms are small.  The deviations in the values obtained 
using the various expressions is within the numerical error in the data.

\begin{figure}
\begin{center}
\includegraphics[scale=0.37]{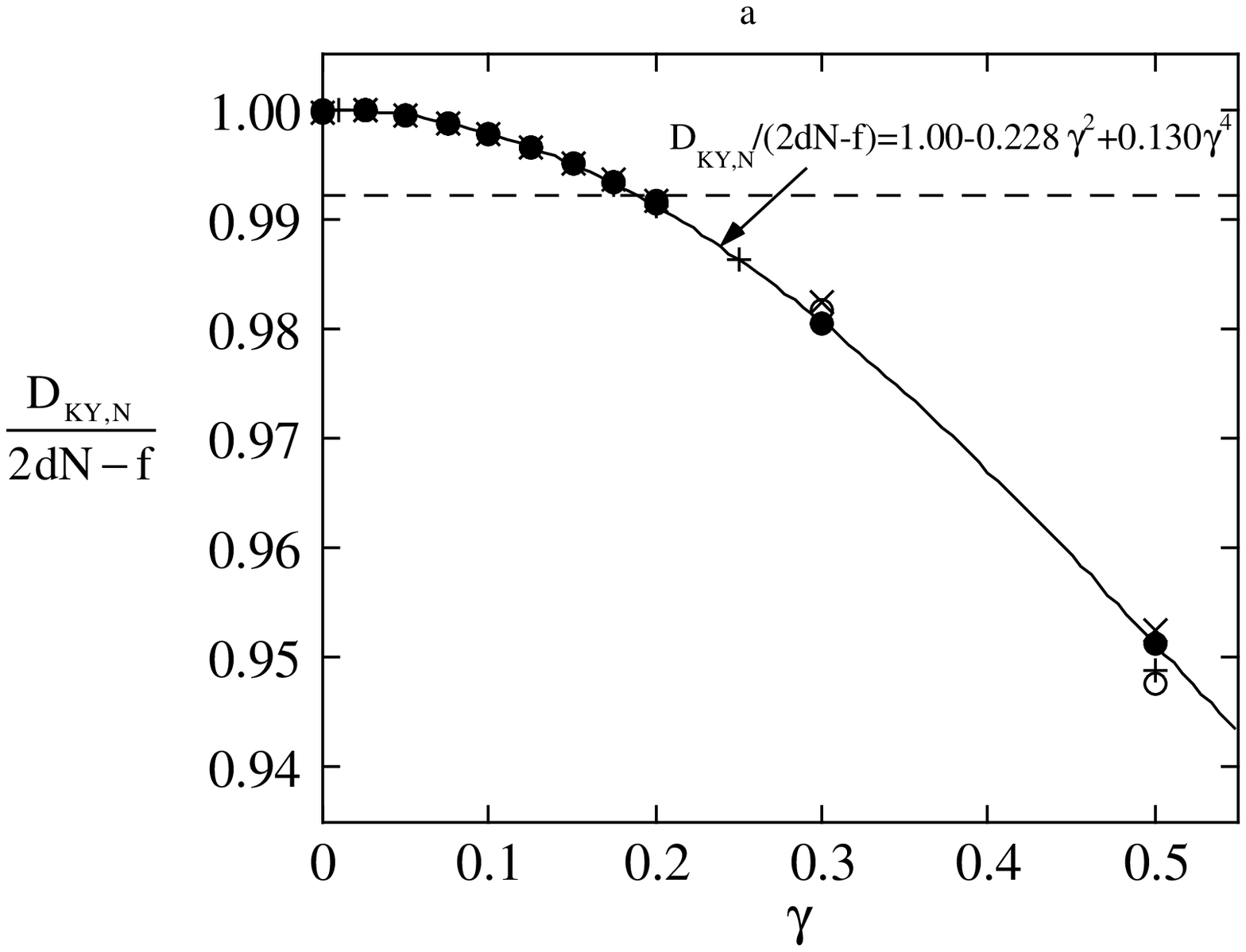}\
\hspace{2cm}\includegraphics[scale=0.37]{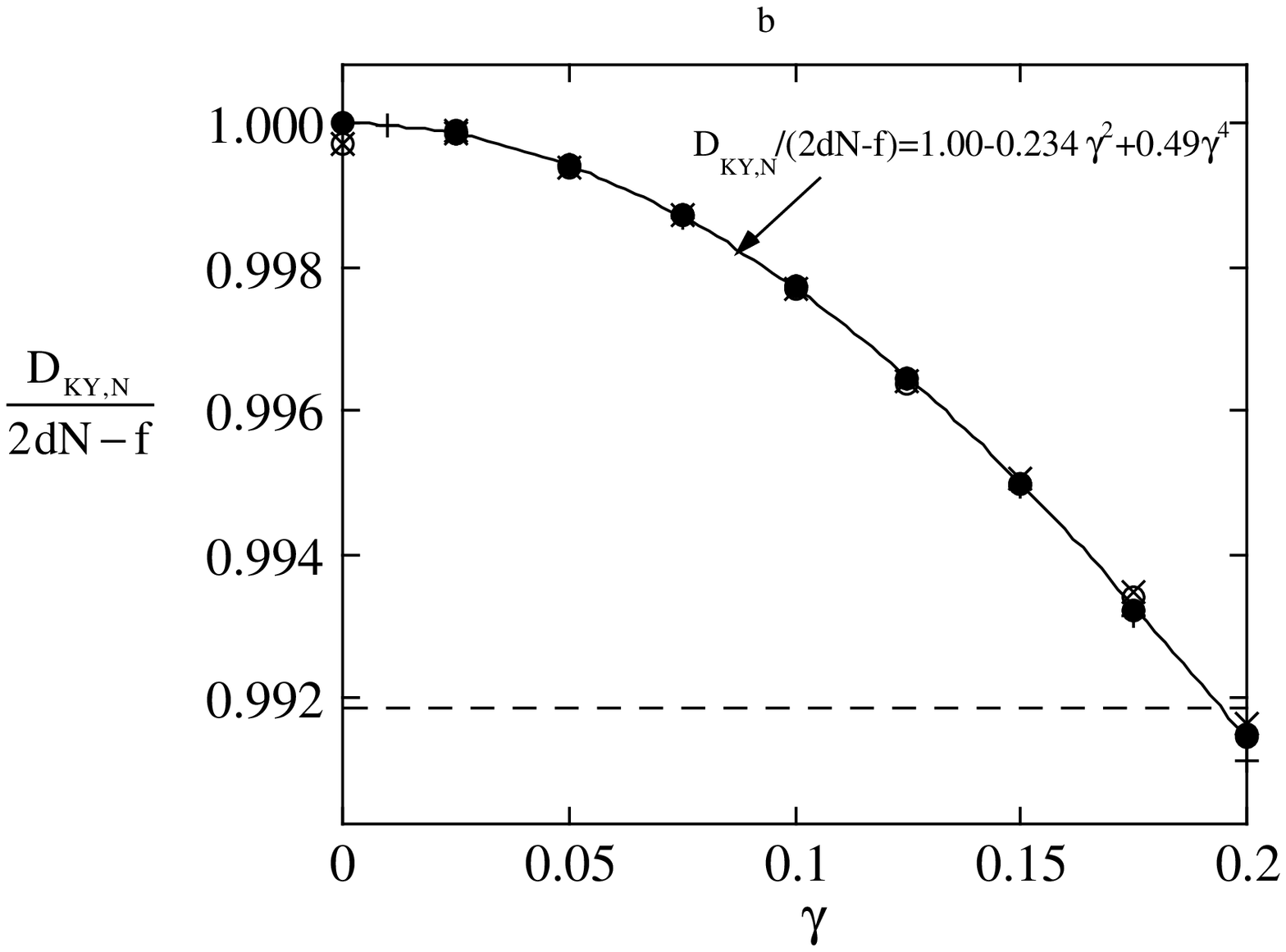}
\caption{Comparison of various expressions for the 
KY--dimension per effective degree of freedom,  
$D_{KY,N}(\gamma)/(2dN-f)$ for sheared systems of $N=32$ particles in $d=2$ 
plotted as a function of the shear rate $\gamma$ up to a strain rate of a) 
$\gamma=0.5$ and b) $\gamma=0.2$.  Plotted are: $D_{KY,N}(\gamma)/(2dN-f)$ 
from its definition Eq.  (\ref{D_KY}) (filled circle); from Eq.  
(\ref{sllod2}) (plus sign); from Eq.  (\ref{sllod3}) \ (crosses); and from 
Eq.  (\ref{sllod4}) (open circle).  The solid line is a fit to Eq.  
(\ref{D_KY}) with a fourth order polynomial even in $\gamma$, and the 
dashed line identifies a phase space contraction of unity.  We emphasize 
that in Eq.  (\ref{sllod2}) the measured $\eta$, deduced independently from 
the constitutive equation, has been used in the computation of 
$D_{KY,N}(\gamma)$.}
\label{Fig4}
\end{center}
\end{figure}

We observe here that $D_{KY,N}$ of Eq. (\ref{D_KY}) is well 
approximated by a fourth 
order even polynomial in $\gamma$ for fields up to $\gamma \sim 
0.5 $ and that at $\gamma \sim 0.3$ the terms of $O(\gamma^4)$ are just 
$5\%$ of the terms of order $O(\gamma^2)$,
while the phase space dimension reduction is clearly greater than unity.  
We note that the quadratic dependence of $D_{KY,N}$ on $\gamma$ implies a 
linear dependence of the shear stress on $\gamma$,\cites{TSE99} so the 
linear regime for $N=32$ extends beyond strain rates where the phase space dimension 
reduction is smaller than one, which runs (for $N=32$) till approximately 
$\gamma=0.175$.  At $\gamma=0.3$, $\eta(\gamma)$ is just $8\%$ smaller than 
$\eta(0)$.

 For completeness, in Figure 4b, we expand Figure 4a in the regime where 
 the phase space reduction is smaller than unity.  In this regime, Eqs.  
 (\ref{KY_e}), (\ref{CPR2}) and (\ref{CPR3}) are expected to be  
 valid for the calculation of $D_{KY,N}$ and any deviations of the data 
 from the value calculated using (\ref{D_KY}) are due to the limited
 numerical precision of the independent calculations of the viscosity and
 the Lyapunov exponents, $O(F_e^4)$ corrections, 
 or to the assumption that the WCPR is obeyed.  The 
 numerical results indicate that the numerical error gives the 
 most significant contribution to the deviations observed.  The WCPR
  is found to be valid at least to within numerical error for 
 the state points considered.  This is consistent with previous 
 work.\cites{SEI98}

Using the fourth order fit to $D_{KY,N}$,  
mentioned above, $\ie$ $D_{KY,N}(\gamma)=2dN-f+{\textstyle{1 \over 
2}}D^{''}_{KY,N} \gamma ^2+{\textstyle{1 \over{4}}}D_{KY,N}^{''''}\gamma^4$ 
for the shear rate range $[-0.2,+0.2]$, (\cf figure 4b), one can calculate 
the zero strain rate or Newtonian shear viscosity from 
$\eta=D_{KY,N}^{''}$$\lambda _{max} nk_BT/2N$.  Here $D_{KY,N}^{''}$ and 
$D^{''''}_{KY,N}$ are the second and fourth derivatives of 
$D_{KY,N}(\gamma)$, respectively, with respect to $\gamma$, taken at 
$\gamma=0$. This leads to a value of $\eta=2.54\pm 0.07$, which agrees 
very well within statistical uncertainties with the Newtonian viscosity 
directly measured in the simulation and calculated from the defining 
constitutive relation $\eta (\gamma )\equiv -P_{xy}(\gamma )/\gamma $, viz.  
$\eta=2.52\pm 0.05$ (see the + points in figure 4b).  Note that although 
the precision of $D_{KY,N}$ is high (less than $0.03\%$ statistical error),
the viscosity is related to the phase space contraction $(2dN-f-D_{KY,N})$ 
and thus calculation of the viscosity from the phase space dimension
reduction involves, at small fields, a very small difference between two 
large numbers, resulting in a larger relative error in the viscosity.  
Furthermore since the dimensional contraction is a quadratic function of 
the shear rate, it results in a more difficult calculation of 
the viscosity from the phase space dimension reduction 
than the constitutive equation which is linear in the shear 
rate at low strain rates.

\begin{figure}
\begin{center}
\includegraphics[scale=0.37]{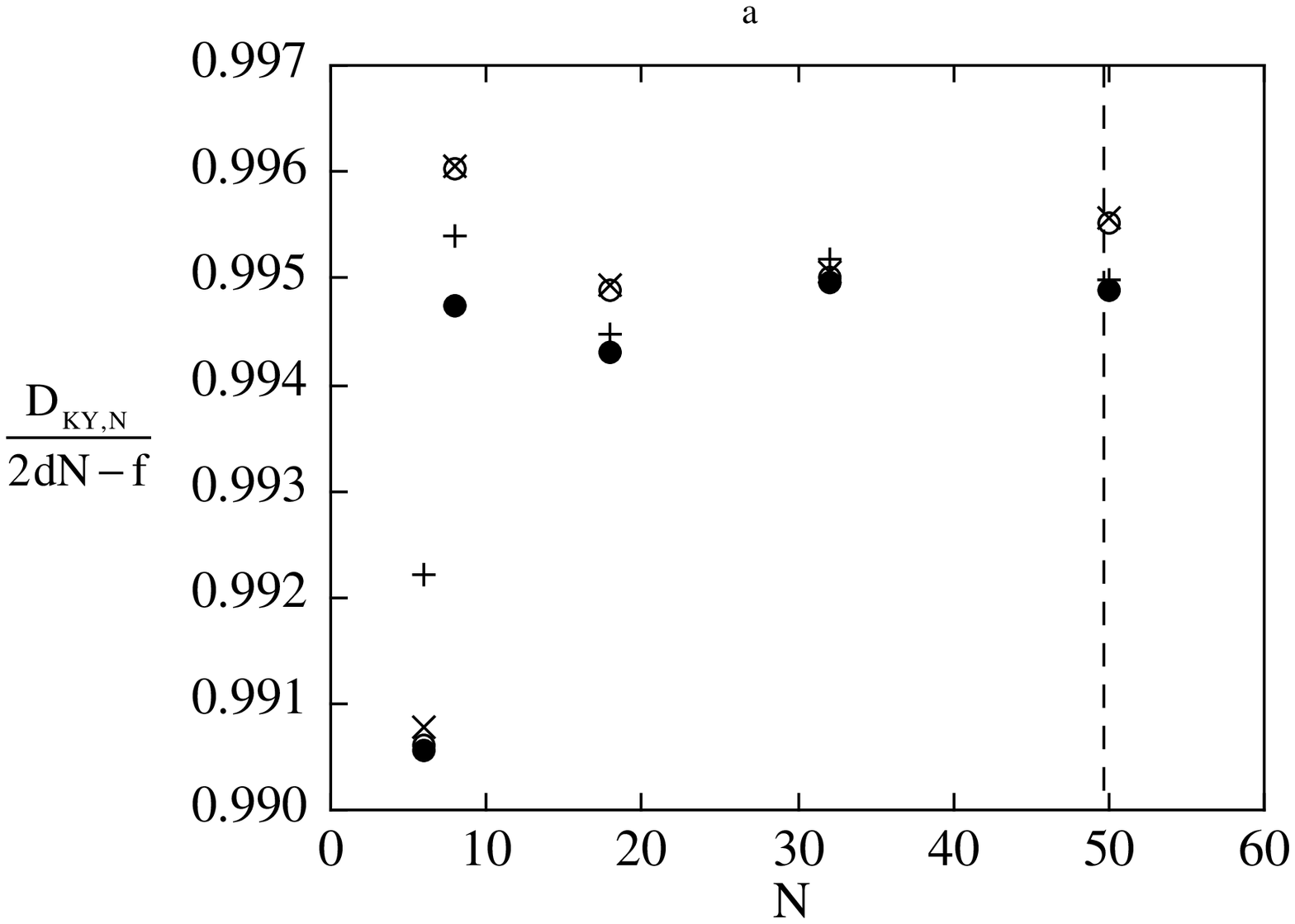}\
\hspace{2cm}\includegraphics[scale=0.37]{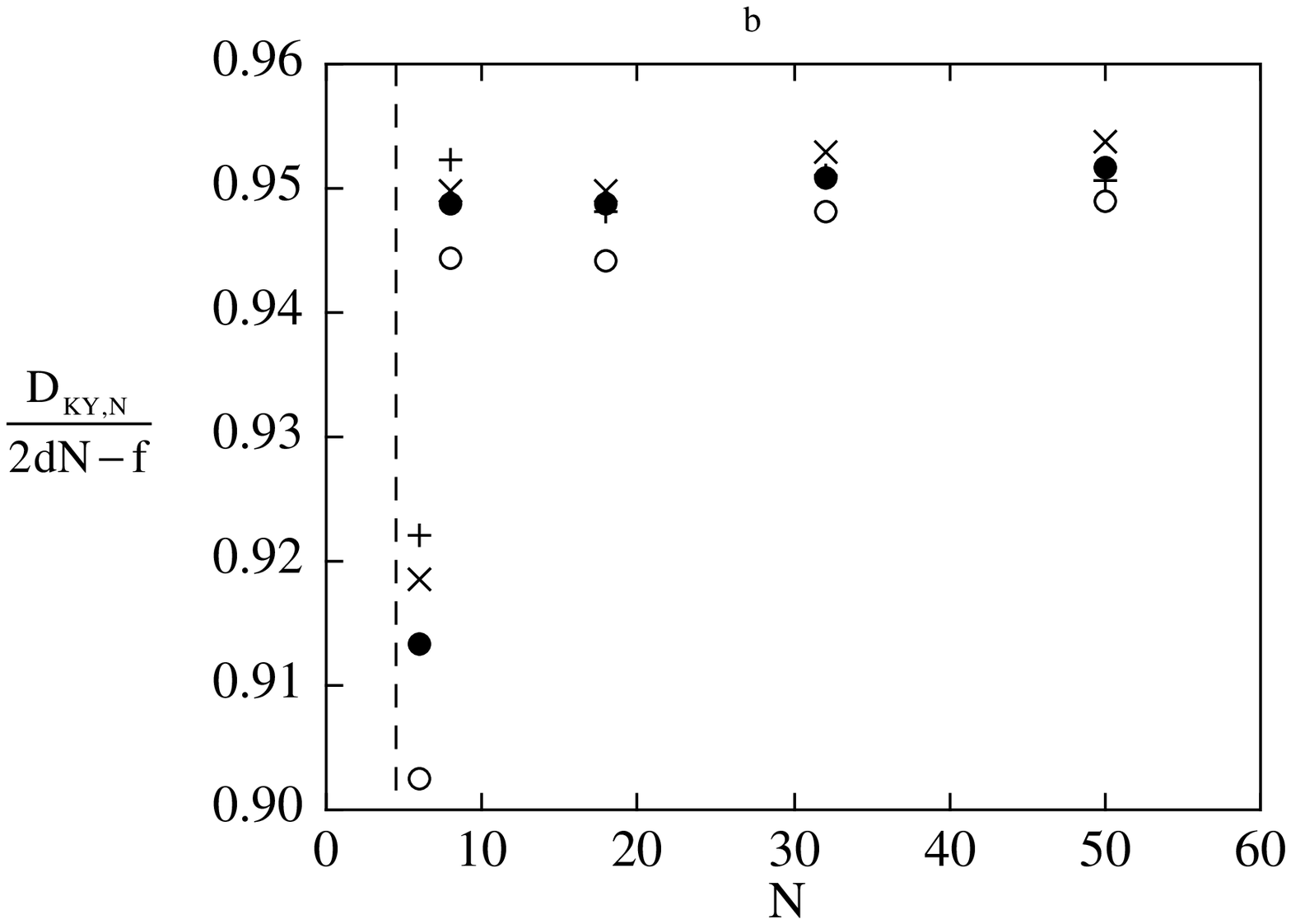}
\caption{Comparison of various expressions for the 
KY--dimension per effective degree of freedom, 
$D_{KY,N}(\gamma)/(2dN-f)$ for sheared systems of WCA particles in $d=2$ at 
a strain rate of a) $0.15$ and b) $0.5$, plotted as a function of the 
number of particles, $N$.  Plotted are: $D_{KY,N}(\gamma)/(2dN-f)$ from its 
definition Eq.  (\ref{D_KY}) (filled circle); from Eq.  (\ref{sllod2}) 
(plus sign); from Eq.  (\ref{sllod3}) (crosses); and from Eq.  
(\ref{sllod4}) (open circle).  Eq. (\ref{KY_e}) can be used to approximate
the system size at which as phase space dimension reduction of unity will occur
for any given field. Using this relation, a phase space dimension reduction less 
than unity is expected to occur for system sizes smaller than that 
indicated by the dashed line.}
\label{Fig5}
\end{center}
\end{figure}

Figure 5 compares the values of $D_{KY,N}$ determined from Eqs.  
(\ref{sllod2}), (\ref{sllod3}) and (\ref{sllod4}) with those calculated from 
its definition in Eq.  (\ref{D_KY}), as a function of $N$ for fields of 
$\gamma=0.15$ and $\gamma=0.5$.  In Figure 5a the results for the field 
of $\gamma=0.15$ are shown, and at this strain rate the dimensional 
contraction will be less than unity for systems size up to approximately 
$N=50$.  Therefore, for $N \stackrel{\scriptstyle<}{\scriptstyle{\sim}} 
50$, Eq.  (\ref{KY_e}), which contains no $O(N^{-1})$ corrections, will be 
valid, as will Eqs.  (\ref{CPR2}) and (\ref{CPR3}) if the WCPR is obeyed.
 All the 
numerical results are consistent with the theory and with the assumption
 that the WCPR is obeyed.\cites{SEI98} At $\gamma=0.15$, the values 
determined using the 
various methods have a maximum difference of $0.2\%$ and are within the
numerical errors at each particle number.  In Figure 5b, the results for a 
strain rate of $\gamma=0.5$ are shown.  In this case, a dimensional 
contraction of less than unity is only obtained for 
$N\stackrel{\scriptstyle<}{\scriptstyle{\sim}}5$.  Again, the deviations 
between the results calculated using Eqs.  (\ref{sllod2})- 
(\ref{sllod4}) and the definition of $D_{KY,N}$ given by Eq.  
(\ref{D_KY}), are small (at most  $1\%$), and within the limits of error for 
all particle numbers considered.  This confirms that the 
coefficients of the 
$O(N^{-1})$ and $O(F_e^4)$ terms are small for this system.

\section{CONCLUSIONS}
\setcounter{equation}{0}

We mention here a few implications of the results presented in this paper.
\begin{itemize}

\item [1.]The extensivity of the phase space reduction for large $N$ and 
small fields is here, to the best of our knowledge, demonstrated for the 
first time, on the basis of the Smoothness Hypothesis of the Lyapunov 
spectrum and the extensivity of the total entropy production.

\item [2.]  The relationships given by Eqs.  (\ref{tran}), 
(\ref{KY_e}) and (\ref{sllod2}) also apply to systems where not all particles 
are thermostatted.  That is, they can be applied to systems where the 
Gaussian thermostat operates on selected particles, say those in the 
boundaries, while the remaining particles evolve under Newtonian dynamics, 
supplemented by a dissipative field.\cites{HP94,AE99} We note that the Eqs.  
(\ref{CPR2}), (\ref{CPR3}), (\ref{sllod3}) and (\ref{sllod4}) can only be
assumed to apply to 
homogeneously thermostatted systems in general, since only for such systems can the 
WCPR be expected to hold.

\item [3.]A simple calculation shows that for a typical case as that of one 
mole of Argon at its triple point, sheared at the rate of 1 Hz, the 
difference of the Kaplan-Yorke
dimension and the phase space dimension $(O(10^{23}))$ is tiny, namely
$\sim 3$.  This follows from Eq.(3.3), which shows that the dimension 
loss, when measured in moles, is equal to the product of the total entropy 
production rate of the system and the reciprocal of the largest Lyapunov 
exponent. Since the largest Lyapunov exponent is controlled by the most 
unstable atomic processes, it is always very small
$\sim$ 1 ps$^{-1}$ whether for atomic, molecular and even polymeric systems.
  
	We note that this smallness of the phase space dimension reduction 
in irreversible processes near equilibrium could well be the reason that
linear Irreversible Thermodynamics provides such a good description
of nonequilibrium systems close to equilibrium.  This is because 
the thermodynamic properties are insensitive to the high order distribution
functions - including the full N-particle distribution function of the 
entire system - since they are determined by a few low order distribution 
functions, which ``do not know'' that the dimension of the steady state 
attractor is only a few dimensions smaller than the $\sim 10^{23}$  
of the phase space of the system.

\item [4.]For the system studied, the $O(N^{-1})$ corrections to 
$D_{KY,N}/(2dN-f)$ that appear in Eqs.  (\ref{sllod2}), (\ref{sllod3}) and 
(\ref{sllod4}), due to the smoothness hypothesis are small (see Figure 5) 
and less than $1.0\%$ even for system sizes of $N=6$.

\item [5.]In conclusion, although the new relations involving $D_{KY,N}$ 
are simple consequences of the Eq.  (\ref{main}), they could nevertheless 
be useful for applications.  In particular the equations give a simple 
expression for the Kaplan--Yorke dimension of the attractor of a class of 
many particle systems close to equilibrium \ie in the regime of linear 
dissipation near equilibrium.

\end{itemize}

\medskip
\section{ACKNOWLEDGEMENTS}
\setcounter{equation}{0} EGDC gratefully acknowledges the hospitality of 
the Research School of Chemistry of the Australian National University as 
well as financial support from the Australian Research Council, the 
Australian Defence Forces Academy and the Engineering Research Program of 
the Office of Basic Energy Sciences of the US Department of Energy under 
Grant No.  DE-FG02-88-ER13847.  DJS and DJE thank the Australian Research 
Council for support of this project.


\begin{thebibliography}{99}
\bibitem{HP88}{H. A. Posch and W. G. Hoover, {\it Phys.  Rev.  A} {\bf 
38}:479 (1988).  We note that the entropy production rate $\sigma$ in this 
paper, as well as in \cite{HP94}, derived from the rate of change of the Gibbs 
entropy which leads to the wrong sign for $\sigma$.  However, this does not 
affect their results.}
\end{thebibliography}
\end{document}